\documentclass[a4paper,12pt]{article}
\usepackage[draft]{hyperref}
\usepackage{multirow}
\usepackage{tabularx}
\usepackage{graphicx, subfigure}
\usepackage{epstopdf}
\usepackage{amsmath}
\usepackage{amssymb}
\makeatletter

\newcommand{\Rmnum}[1]{\expandafter\@slowromancap\romannumeral #1@}
\makeatother

\begin{document}
\title{Periodically patterned columnar thin films as Blazed gratings}

\author{Jhuma Dutta, S. Anantha Ramakrishna, and Akhlesh Lakhtakia}

\maketitle

\begin{abstract}
Periodically patterned columnar thin films (PP-CTFs) were made by evaporating $\mathrm{CaF}_{2}$ and directing the vapor flux obliquely towards lithographically fabricated micrometer/sub-micrometer gratings. The growth of the PP-CTFs was controlled by the deposition rate to form prismatic air cavities within them and they function like blazed diffraction gratings with asymmetric diffraction patterns and diffraction efficiencies upto 52\% in transmission at visible wavelengths. Scalar diffraction theory qualitatively explained the measured diffraction efficiencies.
\end{abstract}

Thin films comprising parallel tilted columns are grown routinely by a variety of physical vapor deposition techniques such as thermal evaporation, electron-beam evaporation, and sputtering \cite{baumeister_book,stf_akhlesh_book}. A collimated vapor flux is generated from a source material and directed towards a planar substrate, the entire process taking place in a low-pressure chamber. The columnar thin films (CTFs) can be grown from a host of source materials such as oxides, fluorides, and even metals. Macroscopically, a CTF is analogous to biaxial crystals and can be used for many different kinds of optical devices \cite{hodgkinson_book}.

Periodic patterning of a substrate at micrometer and sub-micrometer length scales before the deposition of a CTF on it can result in a photonic structure with controllable anisotropy and optical response\cite{nanotech,walsby}. Deposition on a patterned substrate may result in a patterned CTF, because the seed pattern of the substrate shadows out various areas in relation to the arriving collimated vapor flux, thereby preventing deposition in these areas~\cite{nanotech,walsby}. A periodically patterned CTF (PP-CTF) can be expected to have interesting diffraction properties due to a combination of anisotropy and transverse periodicity. Thus, a CTF deposited on a periodic array of rectangular grooves was envisioned to function as a narrowband linear-polarization rejection filter \cite{fiumara}.

In this letter, we present novel PP-CTFs made by evaporating CaF$_2$ on a pre-patterned substrate in such a manner that regular 
prismatic air inclusions are formed, and the PP-CTF functions as a blazed diffraction grating~\cite{blazed_thesis}. Apart from their common use in spectroscopy, there has been recent interest in using blazed gratings for preferential coupling to surface plasmons on a metallic film~\cite{spr1,sambles}, and angularly separating the harmonics generated by intense laser pulses at ultraviolet (UV) wavelengths down to 150 nm~\cite{yeung}. PP-CTFs are readily made with UV-compatible materials like $\mathrm{CaF}_{2}$ and $\mathrm{MgF}_{2}$.  We demonstrate by simple scalar diffraction theory that the asymmetric diffraction from these films is due to the blazing action of the prismatic air cavities.

\begin{figure}[h]
\centerline{\includegraphics[width=8cm]{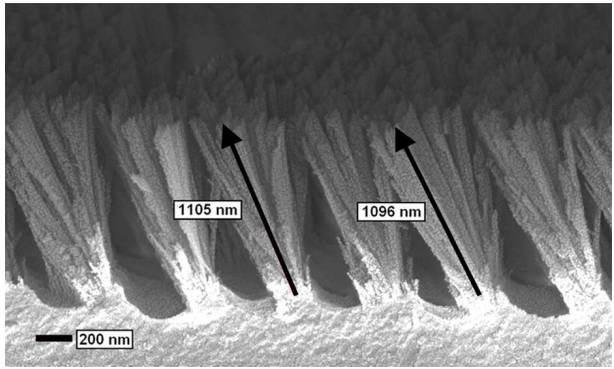}}
\caption{Cross-sectional SEM image of a 1100-nm-thick PP-CTF deposited on a 1D  photoresist grating of 600-nm period. $\mathrm{CaF}_{2}$ was evaporated to deposit the thin film. The prismatic (triangular) air cavities are clearly visible. \label{blz}}
\end{figure}
Laser interference lithography (LIL) was used to fabricate pre-patterned photoresist films that act as periodic seed layers for subsequent deposition of a CTF. A 500-nm-thick film of photoresist (ma-P 1205, Micro-resist Technology) was spin coated on a pre-cleaned smooth glass substrate, baked at 80~$^{\circ}$C, and then exposed to an interference pattern generated by superposing two laser beams from a 473~nm single-longitudinal-mode diode laser. The    photoresist film was then developed to form uniform gratings over areas of 10~mm$^{2}$. One-dimensional gratings with periods of 600 nm and 1125 nm were thus fabricated. Atomic force microscopy (Agilent PicoSPM $\Rmnum{2}$) indicated a modulation of  200-250 nm for these photoresist gratings. CTFs were subsequently deposited by electron-beam evaporation of  $\mathrm{CaF}_{2}$, the collimated vapor being directed at large oblique angles of 10$^{\circ}$-20$^{\circ}$ with respect to the mean substrate plane, and deposition rates of 15-20 {\AA}~s$^{-1}$. The base pressure in the vacuum chamber was $5\times10^{-6}$ mbar that dropped about $1.2\times10^{-5}$ mbar during deposition. As the  vapor flux is incident very obliquely, the ridges of the photoresist grating shadow the valleys of the grating from the collimated flux \cite{nanotech}, resulting in almost no columnar growth in the shadowed regions. Triangular prismatic air cavities are formed within the growing thin film due to the merging of the slightly diverging columnar structures, as revealed by the scanning electron microscope (SEM) image shown in figure~\ref{blz}. The vapor incidence angle and the rate of deposition can be effectively used to change the shape and angles of the prismatic inclusions. We confirmed that the deposited PP-CTFs are reasonably uniform over areas much larger than 10~mm$^{2}$.

\begin{figure}[h]
\centerline{\includegraphics[width=6cm]{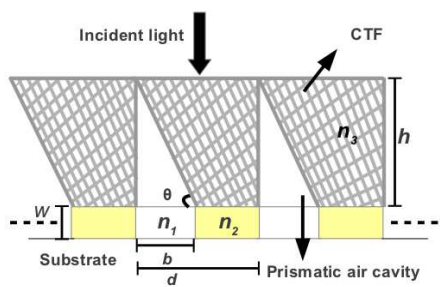}}
\caption{Schematic  of the PP-CTF as a blazed grating containing prismatic air cavities with base $b$, height $h$, and cavity angle  $\theta$. The photoresist grating is made of strips of height $W$ and the period is denoted by $d$.
The refractive indices of the different materials are indicated. 
\label{schematic}}
\end{figure}

These PP-CTFs typically exhibit an asymmetric diffraction pattern in both transmission and reflection, when illuminated normally by a laser, with the diffracted orders on either side having unequal intensities. The asymmetry in diffraction occurs only along the deposition direction. Optical anisotropy of the CTFs by itself can not result in highly asymmetric transmission.
To understand the asymmetry in transmission, we modeled the PP-CTF  as a blazed grating made of an isotropic material of refractive index $n_3$, as depicted schematically in Fig.~\ref{schematic}. The prismatic air cavities give rise to a linear phase shift along the transverse direction in the manner of a blazed grating. Note that, in comparison, a conventional blazed grating has prismatic elements of a high-refractive-index material embedded in a material of low refractive index (air), whereas these PP-CTFs have those elements made of air embedded in a material of refractive index $n_3>1$. 
For CTFs made of CaF$_2$, $n_3$ ranges from 1.43 (bulk $\mathrm{CaF}_{2}$) to 1.1 for highly porous films~\cite{ward_book}.

Assuming a plane wave normally incident on the blazed grating from top and applying the Fresnel--Kirchhoff scalar approximation, we find that the complex amplitude of the transmitted wave in the far zone has two components:
\begin{eqnarray}\nonumber T(k_x) =  \left\{\int_0^b e^{ik_0[(n_1-n_3)x\tan\theta +\uppercase{w}n_1]}
e^{ik_xx}\,dx  \right. \\
 \left. + \int_b^d e^{ik_0(\uppercase{w}n_2 + hn_3)}e^{ik_xx}\,dx\right\} \times T_0  \frac{\sin(\frac{Nk_xd}{2})}{\sin(\frac{k_xd}{2})}\,
\end{eqnarray}
where $T_0$ is  the zeroth-order transmission coefficient, $k_0=2\pi/\lambda$ is the free-space wavenumber, and $N$ is the total number of lines in the grating. The terms within the braces represent the diffraction envelope that is due to diffraction from a single element (i.e., one period) while the last factor arises from interference between light scattered by the periodically placed elements. The linear phase-shifted transmission for the prismatic cavity---the first term of the sum in Eq. (1)---tends to shift the diffraction envelope to one side.

\begin{figure}[t]
\begin{center}
\subfigure{\includegraphics[width=6cm]{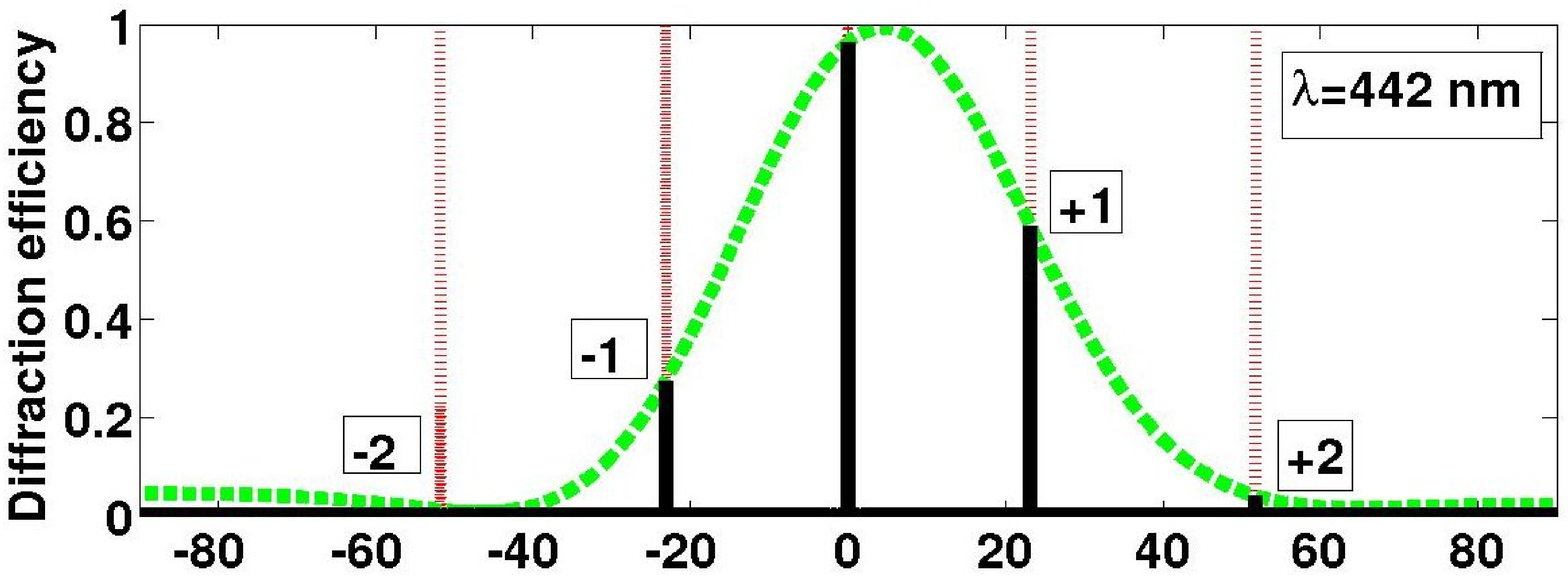}}
\subfigure{\includegraphics[width=6cm]{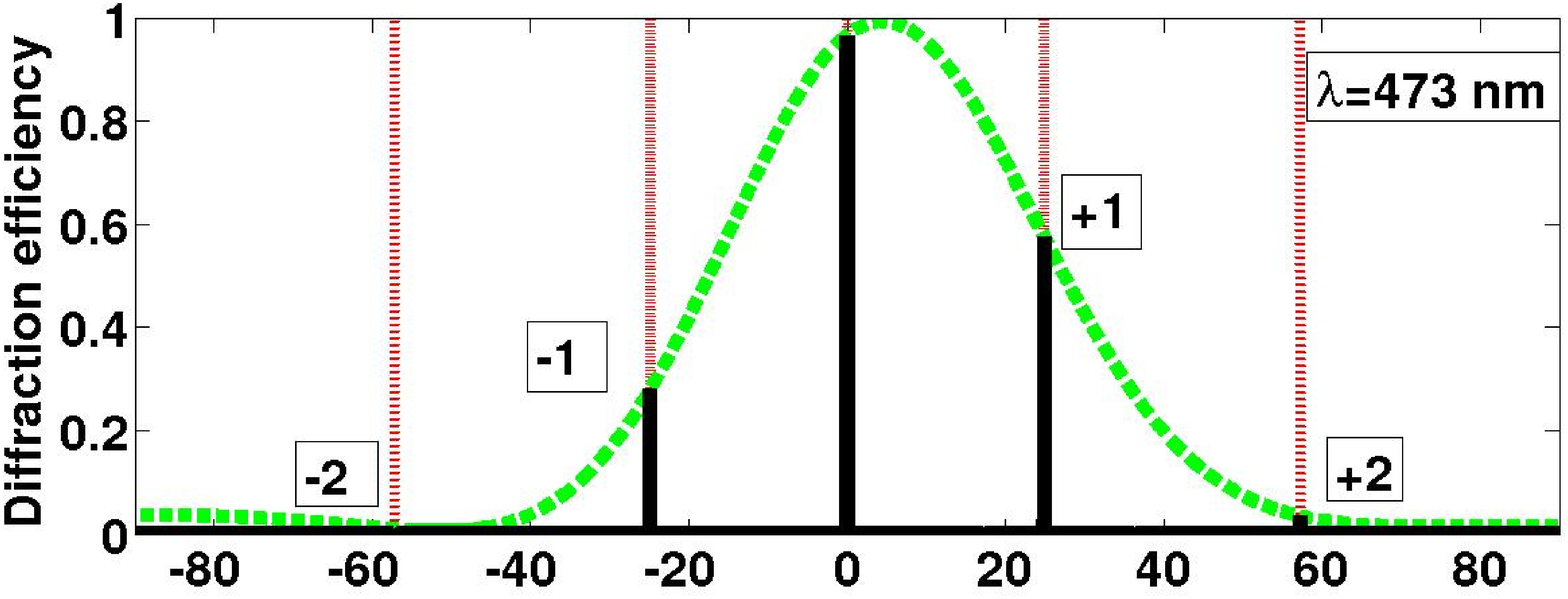}}
\subfigure{\includegraphics[width=6cm]{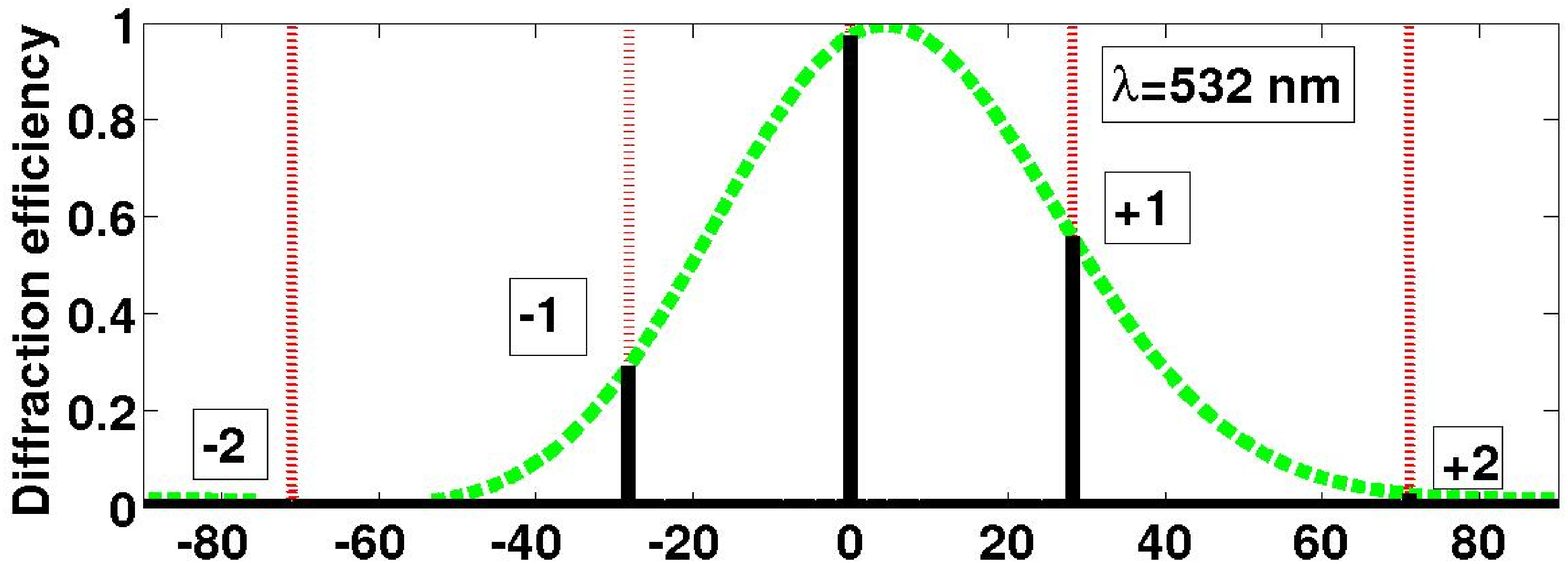}}
\subfigure{\includegraphics[width=6cm]{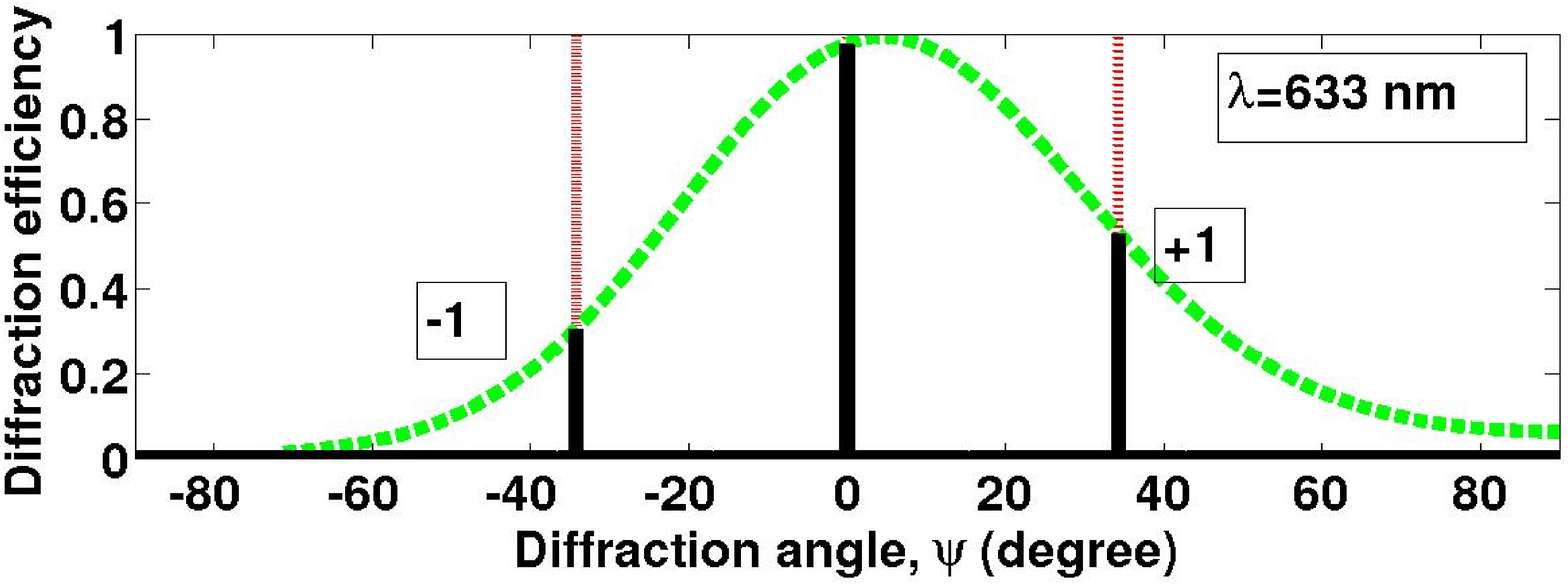}}
\end{center}
\caption{Calculated diffraction efficiency $\vert{T(k_0\sin\psi)/T_0}\vert^2$ in transmission vs. diffraction angle $\psi$ for a blazed grating with period $d=1125$~nm for $\lambda\in\left[442 , 473 , 532 , 633\right]$~nm. Diffraction angles ($\psi$) due to the periodic arrangement are identified by red lines, while the diffraction envelope is a green curve and the diffraction efficiency of each order is represented in black.
\label{subfigure}
}  
\end{figure}
Figure~\ref{subfigure} shows the dependence of the diffraction efficiency $\vert{T(k_0\sin\psi)/T_0}\vert^2$ 
on the diffraction angle $\psi$ calculated for $\lambda\in\left[442 , 473 , 532 , 633\right]$~nm. For these calculations, the following values of the various parameters---obtained by taking averages over five locations on the SEM image of a PP-CTF---were used: $d= 1125$~nm, $h= 1050$~nm, $W=200$~nm, and $\theta$ = 55$^{\circ}$; furthermore, $n_1=1$ (air) and $n_2$ is the refractive index of photoresist\cite{photoresist}. The diffraction envelope maximum is shifted from the zeroth-order direction (i.e., $\psi=0^\circ$) towards the direction of the columnar orientation, indicating  the shift of the transmitted intensity from the zeroth order to positive orders.

For normally incident unpolarized laser beams, diffraction efficiencies in transmission were measured for $\lambda\in\left[442 , 473 , 532 , 633\right]$~nm.
The measured diffraction efficiencies $\vert{T(n2\pi/d)/T_0}\vert^2$ of orders $n\in\left\{\pm1,\pm2\right\}$ are shown in Tables~\ref{table1} and \ref{table2}, for $d = 1125$~nm and $d= 600$~nm, respectively,
along with the diffraction efficiencies predicted by the scalar theory. For the predictions, the refractive index $n_3$ of the PP-CTF (assumed isotropic) was estimated to be $1.12$, which is the volume-averaged refractive index of a 70\% porous film of  $\mathrm{CaF}_{2}$. 

The experimental data and the theoretical predictions agree quite well in Table~\ref{table1}, with the measured diffraction efficiencies consistently lower than their predicted counterparts---most probably, due to the larger reflectivity of the substrate for the diffracted beams at large angles compared to the zeroth-order at normal incidence which is not accounted in Eq. (1). The data in Table~\ref{table2} for the grating with the shorter period, however, reveal a lower level of agreement, although the qualitative trends are maintained. The measured diffraction efficiencies for $\lambda=532$~nm  are much smaller than the predicted values, as expected because the scalar theory becomes inaccurate for smaller features that are comparable to the wavelength. Anyhow, the tables demonstrate that the asymmetric diffraction is the result of the blazing caused by the prismatic cavities. 

We found experimentally that the diffraction efficiencies for incident light polarized perpendicular to the blaze direction are about 1.5 times more than for incident light polarized parallel to that direction. Quantitative predictions will require the use of a fully vectorial diffraction theory that incorporates the anisotropy of the CTFs.

\begin{table}[h]
\caption{Measured and predicted (in parenthesis) diffraction efficiencies for a PP-CTF with $d=1125$~nm.}
\centering
\begin{tabular}{@{\extracolsep{1pt}}lc c rrrr}
\hline\hline
$\lambda$ (nm) &   \multicolumn{4}{c}{Diffraction efficiency (\%)}\\[0.5ex]
\hline
       & $n=1$    & $n=-1$   &  $n=2$   & $n=-2$    \\[0.5ex]
\hline
{442 } & 52(61)   & 24(28)   & 3.4(4.2) & 1.3(1.6)  \\[0.5ex]
\hline
{473 } & 48(59.5) & 26.4(29) & 2(3.5)   & 0.6 (1.23)\\[0.5ex]
\hline
{532 } & 46(57)   & 27(30)   & 1.2(2.8) & nil (0.9) \\[0.5ex]
\hline
{633 } & 44(54.2) & 28(31)   & -        & -         \\[0.5ex]
\hline
\end{tabular}
\label{table1}
\end{table}

\begin{table}[h]
\caption{Same as Table~\ref{table1} except $d=600$~nm.}
\centering
\begin{tabular}{@{\extracolsep{1pt}}c c rr}
\hline\hline
$\lambda$ (nm) & \multicolumn{2}{c}{Diffraction efficiency (\%)}\\[0.5ex]
\hline
                       &  $n=1$   &  $n=-1$    \\[0.5ex]
\hline
{442 }                 &  36 (51)  &  8 (30)    \\[0.5ex]
\hline
{473 }                 &  22 (50)  &  9 (32)    \\[0.5ex]
\hline
{532 }                 &  10 (49)  &  2 (33)    \\[0.5ex]
\hline
\end{tabular}
\label{table2}
\end{table}

The presented fabrication technique can be adapted for UV applications with reactive ion etching to make PP-CTFs on UV transparent substrates for use as transmission gratings with high groove densities and large areas that have certain advantages, particularly when the zeroth order beam is required to go straight through~\cite{Heilmann}.  The oblique angle deposition technique is suited for inexpensive mass manufacture and can work upto much smaller feature sizes down to 150 nm~\cite{walsby}. Another salient point is that the intensities of the diffracted orders depend quite sensitively on the effective permittivity tensor of the CTF which, in turn, depends on the porosity of the film. Measurement of the diffraction efficiencies may provide a way to quantize the CTF porosity, which is otherwise accessible only from gravimetric analysis. 

In conclusion, we have presented periodically patterned columnar thin films made by evaporating $\mathrm{CaF}_{2}$ and directing the resultant collimated vapor flux obliquely towards micrometer/sub-micrometer gratings fabricated by LIL. The formation of prismatic air cavities in the PP-CTF is responsible for highly asymmetric diffraction patterns. A simple model considering the prismatic air cavities within the PP-CTF as blazing elements of a diffraction grating and scalar diffraction theory was successful in capturing the essential features of the measured diffraction efficiencies.  
 
The authors thank H. Wanare (IIT Kanpur) for discussions. JD and SAR acknowledge funding by the CSIR, India under grant no 03(1161)/10/EMR-$\Rmnum{2}$;  JD thanks UGC, India for a fellowship; and AL thanks the Binder Endowment at Penn State for partial financial support.

\end{document}